\def\que#1#2{\displaystyle\frac{#1}{#2}}

\let\Om=\Omega

\let\ov=\overline

\documentclass[pra,twocolumn]{revtex4}

\usepackage{amsfonts,amssymb,bm,graphicx}

\begin{document}

\title{An  analysis of the big-bang theory according to
classical physics}

\author{R. Alvargonz\'alez and L. S. Soto}

\affiliation{Facultad de F\'{\i}sica,  Universidad Complutense,
28040 Madrid, Spain}

\date{\today}

\begin{abstract}
This paper collects a consistent body of information on the
observable Universe, from which an estimate of the total mass of
the Universe is calculated as a function of the angle whose vertex
is at the center of the Universe, and whose extremities stand on
the Earth and on the limits of the horizon of visibility. This
result leads to an analysis of the dynamics of the Big-Bang,
taking into account the limitations imposed by the Schwarzschild
radius, $R_S$. Where if $R_0$ is the radius of the incipient
Universe when the formation of elementary particles has just
finished, the value of the quotient $R_0/R_S$ determines its
subsequent evolution.  An important conclusion from this concerns
the expansion of the Universe; all signs point to its being
destined to expand indefinitely.
\end{abstract}

\maketitle

\section{Preliminary considerations}

This paper uses the $(e,m_e,c)$ system of units in which the basic
nits are the quantum of electric charge, electron mass and the
speed of light.

According to the theory of the Big-Bang, the history of the
Universe begins with an unimaginably great outburst of photons
of very high energy around a point $\omega$, after which time
begins to elapse. The formation of heavy particles at once begins,
and ends when the temperature of the Universe has fallen to below
the required level, and the radius of the Universe has attained
the dimension $R=R_0$.

\begin{figure}[h]
\centering
\resizebox{0.70\columnwidth}{!}{\includegraphics{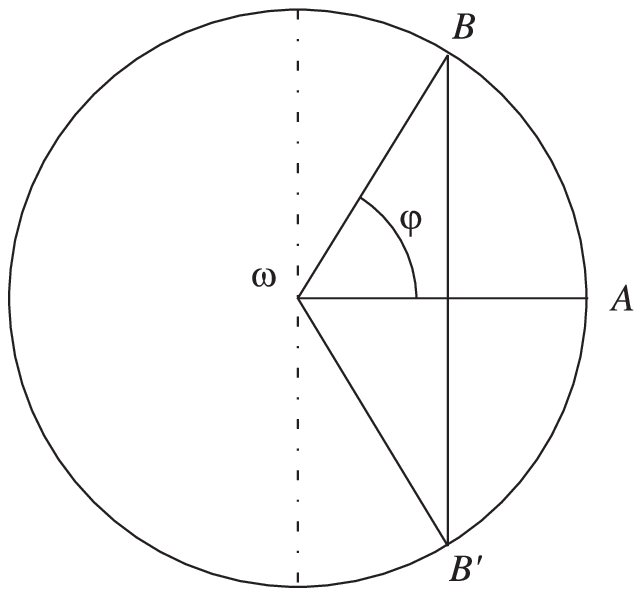}}
\caption{ }
\end{figure}

The dynamics of the Big-Bang thus appears to be the result of the
joint action of the kinetic energies of the particles which
emerge at a velocity close to $c$ at a distance $R_x$ from
$\omega$, and the attraction of the mass of the Universe $M_0$,
which is exerted from $\omega$, the center of its mass, reasonably
assuming a uniform expansion, causing matter to be distributed as
spherical surfaces of radius $R$. Figure 1 shows a diametric section
on which Point $A$ indicates our observatory on Earth, Point $B$
represents the limit of the visibility from $A$, so that the
velocity of growth of $\ov{AB}$ is $c$, and Point $\omega$ is
the centre of the Universe, perhaps the starting point of the
Big-Bang, perhaps the center of what was emerged from the Big-Bang.

The relation between the volume of the observable Universe,
$V_\varphi$, and that of the totality of the Universe, $V_0$,
is given by the relation between the area of the sphere and the
spherical segment $BAB^\prime$. In other words:
$$
\que{V_0}{V_\varphi} = \que{4\pi R^2}{2\pi R^2(1-\cos\varphi)} =
\que2{1-\cos\varphi} .
$$
This relation holds for $\varphi\leq\pi$;  for  $\pi < \varphi < 2\pi$,
it would be $V_0/V_\varphi = 1/(2 + \cos \varphi)$.

Just after the Big-Bang, the radius of the Universe was growing at
speed $c$. As the speed of the growth of $R$ decreased, the value of $
\varphi$ must have gone on increasing, so that the speed at which $B$
moves away from $A$ would continue to be $c$.

\section{Calculation of the mass of the Universe}

The information used in this paper comes mostly from the third
edition of ``Universe", published in 2000. This book contains
(Fig. 26-22) a sky map which is said to show approximately 10 \%
of the visible Universe, and which has permitted an estimate
to be made of the number of galaxies. Roughly $2\times10^6$
galaxies have thus been detected. We can therefore guess that
the number of galaxies in the observable Universe could be of
the order of $2\times10^7$, though this is only a preliminary
estimate.

Also these galaxies are distributed as follows:
\begin{itemize}
\itemsep=-2pt
\item 77 \% spiral galaxies (S) and (SB) of between
$10^9$ and $4\times10^{11} M_\odot$.
\item 20 \% elliptical galaxies (E) of between $10^5$ and $10^{13} M_\odot$.
\item \ 3\% irregular galaxies (Irr) of between $10^7$ and $\times10^{9}M_\odot$.

Where $M_\odot$ is the mass of the Sun ({\em Universe}, p. 646).
\end{itemize}

It is estimated that dispersed interstellar matter
(elementary particles, atoms, molecules and ``dust") is
insignificant in elliptical galaxies and represents between
2.5\% and 25\% of the mass of stars and planets in the
spiral and irregular galaxies.

Keeping in mind the fact that the immense majority of
elliptical galaxies are dwarfs and that the giant elliptical
galaxies are rare, an estimate has been made of the average mass
of the group, assuming that the probability distribution of the
logarithms of their masses follows a Poisson distribution. This
produces an average mass of $10^7M_\odot$ (any future more
exact study would start from the statistical analysis of the
available data --not too awkward a task).

For galaxies (S) and (Irr) a normal distribution has been assumed,
with the result that for $20 \times10^6$ galaxies in question,
we obtain this preliminary estimate:
{\arraycolsep=2pt$$\begin{array}{llcr} \hbox{Spirals} &
3.550855\!\times\!10^{18} M_\odot\\ \\
\hbox{Ellipticals} &
4\!\times\!10^{13}M_\odot\\
\\
\hbox{Irregular} &  3.48\!\times\!10^{14}M_\odot\\
\cline{2-2} &
 3.551243\!\times\!10^{18}M_\odot\end{array}$$}

It may seen absurd to go as far as 7 significant figures in a
preliminary approximation to the mass of spiral galaxies, this has
been done in order to highlight the great preponderance of these
galaxies within the present estimate.

To this sum, $3.5512\times10^{18}M_\odot$, we must add the
intergalactic ``dark matter", scattered through the heart
of the clouds of galaxies, and whose mass is thought to be
of the same order of magnitude as that of the galaxies in
the clouds ({\em Universe}, p\'{a}g. 659). That would mean
that this figure should be doubled, giving the preliminary
result:
\begin{eqnarray}
M_\varphi & = & 7.1025 \times 10^{18}\odot(1.99 \times 10^{33}g/\odot)
\nonumber \\
& \times & 1.097 \times10^{27}m_e/g = 1.55\!\times\!10^{79}m_e ,
\end{eqnarray}
where $M_\varphi$ is the mass of the ``visible Universe", $\varphi$
is the angle $\widehat{A\omega B}$ between the Earth, $\omega$
(origin of the Big-Bang) and $B$, which  is the must distant
observed object and shows a redshift $z=5.34$ ({\em Universe}, p. 597);
$z=\que{\lambda-\lambda_0}\lambda$, where $\lambda$ is
the wavelength of a spectral line observed in a cosmic object,
and $\lambda_0$ is the wavelength of the corresponding line within
the light emitted by an object at rest on Earth.

The equation $\que vc=\que{(z+1)^2-1}{(z+1)^2+1}$ allows us the infer
that the said remotest cosmic object is moving away from us at $v=0.95c$.

The fact that there exists a ``limit of visibility"\ which is
moving away at the speed of ligth shows that the Universe is
expanding. Returning to Fig. 1, and remembering that the length of
the arc $\ov{AB}$ is equal to $R\times\varphi$ ($\varphi$ expressed in
radians), it is clear that the speed at which the length of that
arc increases is \
$\que{d\ov{AB}}{dt}=R\que{d\varphi}{dt}+\varphi\que{dR}{dt}=1\cdot c$

Table I shows as functions of certain values of $\varphi$, the values of the speed
of increase of $R$, $V_R$; the kinetic energy of an electron at rest at $A$,
viewed from a reference system at rest at $\omega$ and whose axis $OX$ coincides
with $\ov{\omega A}$, and the relation $M_0/M_\varphi$ between the mass of the
Universe and that of the observable Universe as estimated above.

\begin{table}
\caption{Values of $V_R/c$; $E_R$ and $M_0$ as functions of $\varphi$.
$\pi$ is the angle $\widehat{A\omega B}$, in radians, where $A$ is the
observatory, $\omega$ the point of origin of the ``Big-Bang" and $B$ is
an object on the ``horizon of visibility" at $1.57 \times 10^{10}$ light
years from $A$. $V_R$ is the velocity of increase of $R$, radius of
the Universe. $E_R$ is the kinetic energy of a particle $m_e$,
moving with velocity $V_R$. $ M_\varphi $ is the mass of the ``observable
Universe", estimated as $1.55 \times 10^{79} m_e$. $M_0$ is the mass of
the Universe, $M_0=2 M_\varphi/(1-\cos\varphi)$.}
\begin{ruledtabular}
\begin{tabular}{||c|c|c|c||}
$\varphi$ (radians) & $V_R/c=1/\varphi$ & $2E_R\,(m_ec^2)=1/\varphi^2$
& $M_0/M_\varphi$ \\
\hline
1.\,00 & 1 & 1 & 4.\,351\\ \hline
1.\,054 & 0.\,949 & 0.\,900 & 3.\,953\\ \hline
1.\,5 & 4/5 & 0.\,640 & 2.\,921\\ \hline
$\pi$/2 & 0.\,6366 & 0.\,405 & 2.\,000\\ \hline
1.\,75 & 4/7 & 0.\,327 & 1.\,697\\ \hline
2.\,00 & 1/2 & 0.\,250 & 1.\,412\\ \hline
2.\,25 & 4/9 & 0.\,197 & 1.\,228\\ \hline
2.\,50 & 2/5 & 0.\,160 & 1.\,110\\ \hline
2.\,75 & 4/11 & 0.\,132 & 1.\,039\\ \hline
3.\,00 & 1/3 & 0.\,111 & 1.\,005\\ \hline
$\pi$ & 0.\,3173 & 0.\,101 & 1.\,000
\end{tabular}
\end{ruledtabular}
\end{table}

This is not the place for deep discussion about the nature of
space in our Universe; the simplest hypotheses are that is
either a three-dimensional Euclidean space, or a three-dimensional
spherical surface. Against the first of these hypotheses, it
can be argued that if it were true, the lines of sight taken
along $\ov{\omega A}$ would imply distances to the ``frontier of
the Universe", which are much shorter than the lines of sight
taken at right angles with $\ov{\omega A}$ (see Fig. 2).
This has not been observed, and would be not the case
if the second hypothesis were true. According to the second
hypothesis the light would move along geodesics within the
three-dimensional surface $x^2+y^2+z^2+u^2=R^2$, which are
great circles of radius $R$, and since we are living inside
it, we would not be in a situation where we could observe
any $4^{\rm th}$ dimension, just as the imaginary ``flatlanders"
living on the two-dimensional surface  $x^2+y^2+z^2=R^2$ could
not perceive the existence of a $3^{\rm rd}$ dimension.

\begin{figure}
\centering
\resizebox{0.50\columnwidth}{!}{\includegraphics{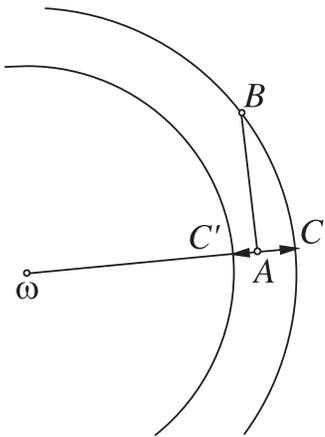}}
\caption{Lines of sight to the frontiers of the Universe in
an Euclidean space. If both $\ov{AC^\prime}$ and $\ov{AC}$ are
greater than the distance from $A$ to the ``horizon of visibility"
no difference would be detected.}
\end{figure}

It has been estimated ({\em Universe}, p. 677) that the value of
the Hubble constant, $H_U$, lies between 60 and 90 km/s. per
megaparsec, i.e., between 18.4 and 27.6 km/s. for every $10^6$
light years. For the lower of these values the distance to the
horizon of visibility would be $1.629\times10^{10}$ light years,
which would thus be the length of the arc $\ov{AB}$ in Fig. 1. For
a value of $H_U=27.6$ km/s. per $10^6$ light years, the length of
the same arc would be only $1.086\times10^{10}$ light years. Table
II shows the lengths of $\ov{AB}$ and $R$ for various values of
$H_U$ and $\varphi$ (in radians).

Nowhere have we mentioned the possibility that $\varphi$ might
be greater than $\pi$ radians. If that were so, we could, after
viewing a cosmic object near to the horizon of visibility, turn
out our telescope in the diametrically opposite direction and
obtain a closer view, but in the other side of that cosmic
object. We must study this possibility by analysis of the
information available from cosmic cartography.
Figure 3 illustrates this possibility, while Table III gives
preliminary values for $v_2/v_1$ and $z_2$ corresponding to
various values of $\varphi$, where $v_2/v_1$ is the relation
between the speed at which the first image is moving away
(the arc $\pi+\varphi$) and that of the second image
(the arc $\pi-\varphi$), and which is given by
$$
\que{v_2}{v_1}= \que{1+\pi/\varphi}{1-\pi/\varphi} . $$

\begin{table}
\caption{Dimensions of $\ov{AB}$ and $R$ for various values of
$H_U$ and $\varphi$. $H_U$ is expressed in km/s per Mpsc and $\ov{AB}$
in $10^{10}$ light years.}
\begin{ruledtabular}
\begin{tabular}{||c|c|c|c|c|c||}
$H_U$ & $\ov{AB}$ & \multicolumn{4}{c||} {$\que
R{10^6 {\rm light\ years}} =
\que{\ov{AB}/\varphi}{10^6\ {\rm light\ years}}$} \\ \hline
\cline{3-6}
 &  & $\varphi=1.25$ & $\varphi=1.50$ & $\varphi=2.00$ & $\varphi=2.50$ \\ \hline
60 & 1.62 & 1.30 & 1.08 & 0.81 & 0.65 \\ \hline
65 & 1.50 & 1.20 & 1.00 & 0.75 & 0.60 \\ \hline
70 & 1.39 & 1.11 & 0.93 & 0.70 & 0.56 \\ \hline
75 & 1.30 & 1.04 & 0.87 & 0.65 & 0.52 \\ \hline
80 & 1.22 & 0.98 & 0.81 & 0.61 & 0.49 \\ \hline
85 & 1.14 & 0.91 & 0.76 & 0.57 & 0.46 \\ \hline
90 & 1.08 & 0.86 & 0.72 & 0.54 & 0.43 \\ \hline
\end{tabular}
\end{ruledtabular}
\end{table}

In turn the value of $z_2$ for the nearest image would be
$$z_2=\left[\que{2\{(z_1+1)^2+1\}}{(z_1+1)^2\left(1-\que{v_2}{v_1}\right)+1+
\que{v_2}{v_1}}-1\right]^{1/2}-1$$

\begin{figure}
\centering
\resizebox{0.70\columnwidth}{!}{\includegraphics{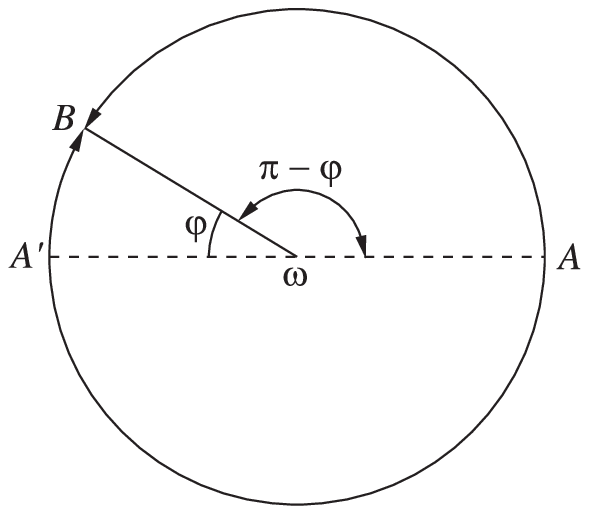}}
\caption{First line of sight $\ov{AA^\prime B}=\pi+\varphi$.
Second line of sight $\ov{AB}=\pi-\varphi$.}
\end{figure}

When we consider the cosmic objects with velocities $v_2$
which correspond to images $(\pi-\varphi)$ of other such objects with $z_1>3.5$,
we must take in account the fact that such images are much more recent, all
the more recent as $\varphi$ becomes larger. It would be necessary to find many
$z_2<\ >z_1$ correspondencies, in diametrically opposite views, in order to
support the hypothesis that the horizon of visibility is at an angle of $\varphi$
greater than $\pi$ radians.

\begin{table}
\caption{Values of $v_2/v_1$ as a function of $\varphi$ and of $z_2$ and
as a function of $\varphi$ and of $z_1$. To each value of
$z_1$ does correspond a value of $v_1/c$.}
\begin{ruledtabular}
\begin{tabular}{||c|c|c|c|c|c|c|c||}
$\varphi$ & & \multicolumn{6}{c||}{Values of $z_1$ and  of $v_1/c$}\\
\cline{3-8}
(radians) & $\que{v_2}{v_1}$ & 6.0 & 5.5 & 5.0 & 4.5 & 4.0 & 3.5\\
\cline{3-8}
& & 0.960 & 0.954 & 0.949 & 0.936 & 0.923 & 0.906 \\ \hline \hline
0.25 & 0.8526 & 2.1650 & 2.1150 & 2.0560 & 1.9840 & 1.8970 & 1.7900 \\ \hline
0.50 & 0.7254 & 1.3640 & 1.3430 & 1.3180 & 1.2870 & 1.2480 & 1.1980 \\ \hline
0.75 & 0.6145 & 0.9690 & 0.9575 & 0.9433 & 0.9256 & 0.9030 & 0.8738 \\ \hline
1.00 & 0.5171 & 0.7238 & 0.7164 & 0.7073 & 0.6959 & 0.6812 & 0.6620 \\ \hline
1.25 & 0.4307 & 0.5524 & 0.5474 & 0.5411 & 0.5332 & 0.5231 & 0.5098 \\ \hline
1.50 & 0.3577 & 0.4242 & 0.4206 & 0.4162 & 0.4106 & 0.4034 & 0.3939 \\ \hline
$\pi$/2 & 1/3   & 0.3933 & 0.3900 & 0.3860 & 0.3809 & 0.3744 & 0.3657\\
\end{tabular}
\end{ruledtabular}
\end{table}

\section{Dynamics of the last particle of mass $m_e$
to emerge from the Big-Bang at a distance $R_0$ from  $\omega$}

Using the $(e,m_e,c)$  system, the mass of the Universe is given
by $M_0m_e$ and the numerical coefficient of the gravitational
constant is $G_e=2.4\times10^{-43}$. When the last particle of
matter was formed as a result of the processes following the Big-Bang,
the formation of pairs of greater mass had ended relatively long ago.
The final addition to the mass derived from the said processes
must have been a pair $e^-$; $e^+$. This electron  of mass $m_e$,
formed at a distance $R_0$ from the center of mass of the Universe,
$\omega$, and moving away from it at a speed very close to
$c$, must have had a kinetic energy in ($R_0$, $t_0$) very close
to $\que12m_ec^2$, being $t_0$ the time elapsed between the
Big-Bang and the formation of that final pair. After a period $t$
had elapse since $t_0$, its distance from $\omega$ must have been
$R_0+R_t$, its velocity $v_t<c$, and its kinetic energy
$E_t=\que12m_e(v_t)^2<\que12m_ec^2$, since the particle in
question was at all times subjected to the attraction of the mass
of the Universe, $M_0m_e$, situated at a distance $R_0+R_t=R$ from
it. We are trying to determine:
\begin{itemize}
\item $E_t$ as a function of $R_t$.
\item The value of $R_t$, when after a determined
lapse $t_\Om$, both $E_t$ and $v_t$ would become null and
the particle begins a journey towards $\omega$.
\item $v_t$ as a function of $R_t$.
\item $R_t$ as a function of $t$.
\end{itemize}

\subsection{Analysis of the evolution of the kinetic energy of the
particle}

The evolution of $E_t$ is given by
$$
E_t= \que12 m_ec^2-\int^{R}_{R_0} \que{(M_0m_e)m_edR}{(Rl_e)^2}
\times2.4\times  10^{-43}\que{l_e^3}{m_et_e^2} ,$$
$l_e=e^2/(m_ec^2)$ and $t_e= e^2/(m_ec^3)$ being the units of
length and time in the $(e,m_e,c)$ system of units, and
$R=(R_0+R_t)$. Obviously $l_e/t_e=c$.

By multiplying out and taking $K_0=M_0 2.4\times10^{-43}$ we
obtain
$$
\que{E_t}{m_ec^2}=\que12-\int^{R_0+R_t}_{R_0}\que{K_0dR}{R^2},
$$
which integrates to
\begin{equation}
\que{E_t}{m_ec^2}=\que12-\left[\que{K_0}{R}\right]^{R_0+R_t}_{R_0}=
\que12-\que{K_0}{R_0}+\que{K_0}{R_0+R_t}.
\end{equation}

The condition for $E_t$ being null is
$$
\que12-\que{K_0R_t}{R_0(R_0+R_t)}=0 .
$$

Remembering that the Schwarzschild radius $R_S$ for the mass
$M_0m_e$ is
\begin{eqnarray*}
R_S & = & \que{2(M_0m_e)G_e}{c^2}=\que{2M_0m_e}{c^2}\cdot2.4\times10^{-43}
\que{l_ec^2}{m_e}\\
& = & 4.8\times10^{-43} M_0 l_e,
\end{eqnarray*}
that is $2K_0l_e$, the
condition (2) can be also written as
$$\que12-\que{\que12R_SR_t}{R_0(R_0+R_t)}=0;$$
whence:
\begin{equation}
R_t = \que{R_0^2}{R_S-R_0}\geq 0 .
\end{equation}

\begin{itemize}
\item When $R_0=R_S$, $E_t$ is null for $R_t \rightarrow \infty$, and we have an
unlimited expansion with a kinetic energy tending towards zero, and therefore
a velocity also tending towards zero.
\item When $0<R_0<R_S$, $R_t$ increases towards a maximum value
$R_t=\que{R_0^2}{R_S-R_0}$, and when it reaches this value, it begins to
decrease, which means that the particle returns towards $\omega$.
\item When $R_0>R_S$, $E_t$ can never get null; $\que{K_0}{R_0}$ in (2) is
always less than 1/2, and $R_S<R_0$. The particle under consideration
continues to recede from $\omega$, but its kinetic energy does not tend towards
zero as $t$ increases indefinitely. It tends towards
$\que12\left(1-\que{R_S}{R_0}\right)m_ec^2$, to which corresponds the velocity
$v=\left(1-\que{R_S}{R_0}\right)^{1/2} c$.
\end{itemize}

\subsection{Analysis of the evolution of the velocity of the particle}

The evolution of the velocity of the particle, $v_t$, as a function of
time starts from $R_0$, $t_0$; where and when $v_t\cong c$, i. e.
$1\cdot c$ in the system $(e,m_e,c)$. From then on, its velocity
decreases, because the particle of mass $m_e$ is subjected to the
attraction of the mass $M_0m_e$, which is distant $Rl_e=(R_0+R_t)l_e$
from it, and which exerts  on it the force
$$
f(t)=\que{K_0}{(R_0+R_t)^2}\que{m_el_e}{(t_e)^2}=\que{R_S}{2(R_0+R_t)^2}
\que{m_el_e}{(t_e)^2}
$$

This force, applied to the particle of mass $m_e$, determines that this
particle, emerging at $(t_0,R_0)$, and moving away from $\omega$ at a velocity
very close to $c$, suffers a deceleration of:
$$-a(t)=-\que{R_S}{2(R_0+R_t)^2}\que{l_e}{t_e^2},$$
for every $dt$. We can therefore write:
\begin{equation}
v(t)=\que{dR}{dt}=c-\int^t_{t_0}\que{R_S}{2(R_0+R_t)^2}d t .
\end{equation}

From this we derive:
$$
\que{d^2R}{(dt)^2}=-\que{R_S}{2(R_0+R_t)^2}
$$

By next multiplying both these terms by $2dR$ we obtain:
$$2dR\cdot\que{d^2R}{dt^2}=\que d{dt}\left(\que{dR}{dt}\right)^2=
(-1)\cdot\que{R_S\cdot 2dR}{2R^2},$$
because $R_0+R_t=R$.

By integrating we have
$\left(\que{dR}{dt}\right)^2=(-1)\cdot\que{-R_S}{R}+K_0$, whence:
$$\que{dR}{dt}=\left (\que{R_S}{R_0+R_t}+K_0\right)^{1/2} .$$

For $t=t_0$, $R_t=0$; $\que{dR}{dt}=1$ \ and by introducing these
values in the precedent equation, we obtain $K_0=1-\que{R_S}{R_0}$,
and finally:
\begin{equation}
\que{dR}{dt}=\left (1-\que{R_S}{R_0}+\que{R_S}{R_0+R_t}\right)^{1/2} .
\end{equation}

The condition for $\que{dR}{dt}$ being zero is:
$$1-\que{R_S}{R_0}+\que{R_S}{R_0+R_t}=0,$$
whence
$$R_t=\que{(R_0)^2}{R_S-R_0}=\que{R_0}{\que{R_S}{R_0}-1} ,$$
which is identical to (3).

The relation (5) can be written
$\que{dR}{dt}=\left(1-\que{R_S}{R_0}+\que{R_S}R\right)^{1/2}$. Therefore
$1-\que{R_S}{R_0}+\que{R_S}R\geq0$
\begin{itemize}
\item For $R_S/R_0\geq1$; \ $R_S/R_0=1+x$; \ $R_S/R\geq x$.
If this condition is fulfilled, $R$ grows up to
$$R=R_0+\que{R_0}{R_S/R_0-1}=\que{R_S}{R_S/R_0-1},$$
and thereafter decreases towards $\omega$. When by decreasing
reaches $R_0$, $dR/dt$ will be equal to $-c$, which is a boundary
that, out of black hole conditions, can not be trespassed.

\item For $R_S/R_0\leq 1$, \ the condition for the annulation of
$dR/dt$ cannot be fulfilled because the condition (3) would lead
to negative values of $R_t$. When $R_S/R_0=1-x$, $ x \rightarrow 0$,
the value of  $dR/dt\rightarrow0$ for $t \rightarrow \infty$, but
never can be equal to zero, as in the aforementioned case when
$R_S/R_0=1+x$, \ $x\rightarrow 0$. In all the other cases
$R_S/R_0<1$ means that, when both $t$ and $R\rightarrow\infty$
\begin{equation}
\que{dR}{dt}\rightarrow\left(1-\que{R_S}{R_0}\right)^{1/2} .
\end{equation}
\end{itemize}

\subsection{Analysis of the evolution of $R$}

In the last analysis we obtained at (5)
$$
\que{dR}{dt}=\left(1-\que{R_S}{R_0}+\que{R_S}{R_0+R_t}\right)^{1/2}=
\left(1-\que{R_S}{R_0}+\que{R_S}R\right)^{1/2},
$$
whence we obtain:
\begin{equation}
\que{dR}{\left(1-\que{R_S}{R_0}+\que{R_S}R\right)^{1/2}}=dt .
\end{equation}

Integral (171), on page 409 of the 18$^{\rm th}$ edition  of the
``CRC Ma\-the\-ma\-ti\-cal Tables", states that
\begin{equation}
\int\que{x^2dx}{\sqrt{x^2\pm a^2}}=
\que x2(x^2\pm a^2)^{1/2}\mp\que{a^2}2\log [x+(x^2+a^2)^{1/2} ] .
\end{equation}
If we make $R=x^2$, \ $\que{R_S}{(1-R_S/R_0)}=a^2$, equation (7) becomes
$$\que{2}{(1-R_S/R_0)^{1/2}}\cdot\que{x^2dx}{(x^2+a^2)^{1/2}}=dt$$

If we apply solution (8) to the first term of this equation we obtain
\begin{eqnarray*}
& \que1{(1-R_S/R_0)^{1/2}} \left\{ R^{1/2}\left ( R+\que{R_S}{1-R_S/R_0}\right)^{1/2}
\right . & \\
& \left.- \que{R_S}{(1-R_S/R_0)^{1/2}} \log  \left [R^{1/2}+
\left(R+\que{R_S}{1-R_S/R_0}\right)^{1/2} \right ] \right \} & \\
& =t+K_2 .&
\end{eqnarray*}

For $t=0$; \ $R_t=0$ \ and \ $\que{dR}{dt}$ must equal 1, which is in fact
what happens in (5). From the last equation we obtain
$$K_2=\que{R_0}{1-\que{R_S}{R_0}}-\que{R_S}{1-\que{R_S}{R_0}}
\log\left [ R_0^{1/2}+\que1{\left(1-\que{R_S}{R_0}\right)^{1/2}} \right ] . $$

By introducing this value of $K_2$ into the same equation we obtain
\begin{eqnarray}
& \que1{\left ( 1- \que{R_S}{R_0}\right)^{1/2}}
\left[ R^{1/2} \left (R + \que{R_S}{1-\que{R_S}{R_0}}\right)^{1/2} -
\que{R_S}{\left(1-\que{R_S}{R_0}\right)^{1/2}} \right . &  \nonumber \\
& \left . \times \log \que{R^{1/2}\left(1-\que{R_S}{R_0}\right)^{1/2}+
\left\{R\left(1-\que{R_S}{R_0}\right)+R_S\right\}^{1/2}}{R_0^{1/2}
\left(1-\que{R_S}{R_0}\right)^{1/2}+R_0^{1/2}}\!\right] = & \nonumber \\
& =t+\que{R_0}{1-\que{R_S}{R_0}} . &
\end{eqnarray}
By making $A=(1-R_S/R_0)$ this equation may be written
\begin{eqnarray}
& R \left(A + \que{R_S}R\right)^{1/2} -
\que{R_S}{A^{1/2}} \log \left(\que R{R_0} \right)^{1/2} & \nonumber \\
& \times \left[ \que{A^{1/2} + \left( A + \que{R_S}R\right)^{1/2}}
{A^{1/2}+1} \right ]=  At + R_0 .
\end{eqnarray}

When considering these equations we must remember that the equivalence
$R_S/(1-R_S/R_0=a^2$ requires that $R_S\leq R_0$. This means that we must
discard those solutions where $R_0<R_S$. As we have seen in the preceding analysis, these solutions
are those which imply that $R$ increases up to a limit
$R_{\rm max}$, after which it would decrease, thus returning the Universe
towards $\omega$. In other words, (8), (9) and (10) imply
the negation of the Big Crunch. This should not be seen as a binding proof; it
derives from the change of variables required to solve an integral. However,
the fact that if the contrary case is true, its solution leads to imaginary
values is surely a strong argument against that.

Equations (9) and (10) show that the function $R(t)$ depends only on the
quotient $R_0/R_S$. We must remember that $R_0$ is the length of $R$ when the
last pair $e^-$; $e^+$ was formed following the Big-Bang, and that $R_S$ is the
Schwarzschild radius for the mass of the Universe.

The value of $dR/dt$ for $t=0$, \ $R=R_0$, \ is 1. When both $t$ and $R$
increase, the value of
$$\que{dR}{dt}=\left(1-\que{R_S}{R_0}+\que{R_S}R\right)^{1/2}$$
diminishes and when both variables tend to $\infty$, \ $dR/dt$ tends to
$(1-R_S/R_0)^{1/2}$. In this case (10) can be written
$$A^{1/2}(R-A^{1/2}t)=\que R{A^{1/2}}\log\left\{\left(\que R{R_0}\right)^{1/2}
\que{2A^{1/2}}{A^{1/2}+1}\right\}+R_0$$

\begin{figure}[h]
\centering
\resizebox{0.70\columnwidth}{!}{\includegraphics{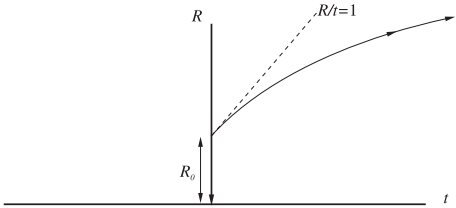}}
\caption{Representation of the function (10). }
\end{figure}

The value for $t\rightarrow0$ of the asymptotic direction $R=A^{1/2}t$ is given
by
\begin{eqnarray*}
& \lim_{t\rightarrow\infty}(R-A^{1/2}t)=\lim\que{R_S}A
\log\left [ \left(\que
R{R_0}\right)^{1/2} \que{2A^{1/2}}{A^{1/2}+1}\right] & \\
& \que{R_0}{A^{1/2}}=\infty . &
\end{eqnarray*}

Therefore we have a parabolic branch parallel to $R=A^{1/2}t$. Knowing this
and the value of $dR/dt$ for the starting point $t=0$, $R=R_0$, we can easily
draw the curve for (10). Only the values for $t>0$, $R>R_0$ are valid (see
Fig.~4).

For $R_S/R_0=1-x$; $x \rightarrow 0$,  $A \rightarrow x$ and (10) can be written
\begin{eqnarray*}
& R \left(x+\que{R_S}R\right)^{1/2}-\que{R_S}{x^{1/2}}\times & \\
& \log \que{x^{1/2}(R/R_0)^{1/2}+\{x(R/R_0)+1-x\}^{1/2}}{x^2+1} & \\
& = xt+R_S(1+x) . &
\end{eqnarray*}
When $x \rightarrow 0$
\begin{eqnarray*}
& \lim_{x \rightarrow 0}\que{R_S}{x^{1/2}}\log\que{x^{1/2}\left(\que
R{R_0}\right)^{1/2}+\left\{x\left(\que
R{R_0}\right)+1-x\right\}^{1/2}}{x^2+1}& \\
& = \lim_{x\rightarrow0}R_S\cdot\que{\log(1-x)^{1/2}}{x^{1/2}}=0. &
\end{eqnarray*}

Therefore, for \ $x\rightarrow0$ \ we have
$R^{1/2}=\que{xt}{R_S^{1/2}}+R_S^{1/2}$.

For each value of $x$ we  have a function $R(t,R_S)$. We know that $x$ is very
little; it tends to zero, but it is not equal to zero, and for each value of
$t$ we have a function like that of Fig. 4, with a value of $dR/dt$ given by
$dR/dt=(x+R_S/R)^{1/2}$, which tends to zero when $t\rightarrow\infty$ and
$R\rightarrow\infty$, in agreement with the analyses developed in 3.1 and 3.2.

In line with (10), it turns out that the value of $dR/dt$, which starts from
being 1, diminishes when both $t$ and $R$ grow greater, and tends to
$dR/dt=(1-R_S/R_0)^{1/2}$, when these varialbles tend to $\infty$. Therefore
the value of
$$R=\int^R_{R_0}\que{dR}{dt}\cdot dt,$$
must be of the same order of magnitude as $t$, and when \ $t\rightarrow\infty$, \
$R\rightarrow\infty$ \ but \ $R<t$.

\begin{center}*\ \ \ *\ \ \ *\end{center}\vskip 16pt

We must now consider the measurement $R_0$ of the radius of the
Universe at the moment of completion of the formation of matter.
The Big-Bang might either have happened exactly at the point
$\omega$, or have burst out as a ``sphere of light"\ of radius $R_x$,
without this making any difference to the eventual evolution of
the Universe. What is really important is that the speed of
formation of the matter existing after the elapse of $t$ since the
happening of the Big-Bang is such that $R_t=R_x+c\cdot t>2M_t\cdot
G/c^2,$ which in the $(e,m_e,c)$ system can be written as
$R_x+c\cdot t>4.8\times10^{-43}M_t\cdot l_e,$ where $M_t$ is the
mass of the Universe (a number of $m_e$), after the elapse of $t$
(a number of $t_e$), and where $R_x+t$ is the radius of the
Universe at that time, a number of $l_e$.

In accordance with the first of these alternatives, $R_x=0$, and the mass of
the Universe after $t$ has elapsed since the start of time must fulfill the
condition
$$M_t<\que{m_e}{4.8\times10^{-43}}\cdot\que{ct}{l_e}$$

For $t=1s.=1.063871\times10^{23}t_e$, the radius of the Universe
would measure $R_1=1.063871\times10^{23}l_e$, for which value the
``mass of Schwarzschild" would be $2.216398\times10^{65}m_e$. A
greater speed of transformation into matter of photons of very
high energy, would result in the formation of a black hole.
Therefore, the process of formation of matter, up to a total
estimated at not less than $1.55\times10^{79}m_e$ (see (1)), must
have lasted for at least
\begin{eqnarray*}
& \que{1.55\times10^{79}m_e}{2.216398\times10^{65}m_e/{\rm s.}}=
6.993\times10^{13}{\rm s} & \\
& = 2.216\times10^6\ {\rm years}.  &
\end{eqnarray*}

This value is a very strich bottom limit, since we must suppose
that the speed of matter formation decreases as the density of
energy decreases, together with the increase in the radius of the
Universe and the disappearance of the photons of very high energy,
which had already been consumed during the previous formation of
matter. This limit obviously corresponds to the Schwarzschild
radius for $1.55\times10^{79}m_e$.

As for the second alternative, in which the Big-Bang starts with
the sudden outburst of a ``sphere of light"\ of radius $R_x$, the
radius of the Universe at the moment of completion of matter
formation, would still be greater than $2.216\times10^6$ light
years, but the time-lapse needed to pass this limit would be
\begin{eqnarray*}
& \que1c(2.216\times10^6\ {\rm light\ years}-R_x\ {\rm light\ years})& \\
& = (2.216\times10^6-R_x)\ {\rm years,} &
\end{eqnarray*}
and the speed of matter formation could have been much greater than
$2.216\times10^{65}\ m_e/{\rm s.}$

According to {\em Universe} (p. 736):
\begin{quote}
``Thus, as soon as the temperature of the radiation field falls
below $1.09\times10^{13}K^0$, photon-antiproton pairs can no
longer be created. Each kind of particle has its own threshold
temperature. For example, electrons and positrons have masses 2000
times smaller than protons and antiprotons. Thus the threshold
temperature for the creation of electron-positron pairs is about
1/2000 that for proton-antiproton pairs. Con\-se\-quen\-tly, when
the radiation temperature falls below $5.93\times10^9K^0$, the
reation $\gamma + \gamma =e^+ + e^-$ that creates electron-positron
pairs can no longer take place".
\end{quote}

From what immediately follows this lines from {\em Universe}, it
can be inferred that the current hypotheses on the evolution of
the Universe includes the supposition that the formation of matter
was completed before 2 seconds has passed since the start of time,
which implies that the Big-Bang happened as a sudden outburst of a
``sphere of light", of radius greater that $2.216\times10^6$ light
years.

This reasonings that have been developed for
$R_S=2.216\times10^6$ light years, which correspond to $\varphi=\pi$
radians, $M_0=M_\varphi$, would be equally repeated for the any
value of $M_0=\que2{1-\cos\varphi}$; \ $1<\varphi<\pi$. On Table I we can
see that $M_0$ may variate between $1.00M_\varphi$ and $4.351M_\varphi$.

\begin{center}*\ \ \ *\ \ \ *\end{center}\vskip 16pt

Let $R_0$ be the radius of the Universe at the completion of
formation of matter in $t=t_0$, and $R_1$ and $R_2$ the distances
to $\omega$ of 2 particles which were formed during $t<t_0$. If the
order of magnitude of $t_0$ is $2{\rm
s.}=2.127742\times10^{23}t_e$ and that of $R_0$ is greater than
$7.439663\times10^{36}l_e$, the values of $R_1$ and $R_2$ would be
expressed for the whole radius $\ov{\omega R_0}$ in very huge
numbers, while the time elapsed between $t=0$ and $t=t_0$ will
also produce huge figures for the practical whole of that time.

\begin{figure}[h]
\centering
\resizebox{0.70\columnwidth}{!}{\includegraphics{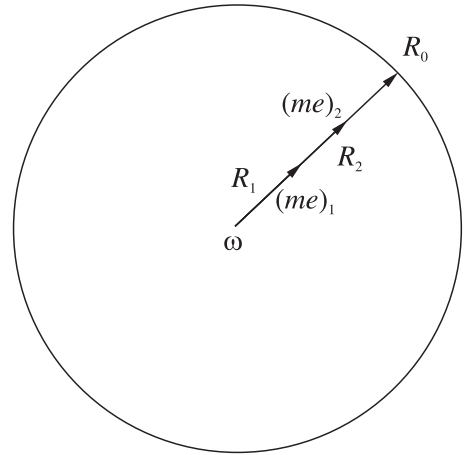}}
\caption{Positions of $(m_e)$, and $(m_e)_2$
at $t=t_0$, \ $R=R_0$.}
\end{figure}

The particle $(m_e)_1$ formed at distance $R_1$ from $\omega$ in $t=t_1$,
is only affected by the attraction of those particles whose distance from
$\omega$ is less that $R_1$. Also, the density of the matter formed between $t=0$
and $t=t_1$ must be uniform over the whole Universe if, as seems logical, we
suppose that the ``sphere of light"\ which arose in the Big-Bang had an equal
density of energy in all its parts. If so, the mass which would attract
$(m_e)_1$, from $\omega$ would be equal to $\que43\pi(R_1)^3\delta$, and the mass
which would attract $(m_e)_2$ would be
$\que43\pi(R_2)^3\delta$. Thus, there corresponds to position $R_x$ the
Schwarzschild radius:
$$R_{S_x}\cdot l_e=\que{8\pi}3\delta(R_x)^32.4\times10^{-43}l_e=
6.4\cdot10^{-43}\pi\delta(R_x)^3l_e$$

By writing $R_x=xR_0$, \ $x<1$, we obtain
$$R_{S_x}=6.4\cdot10^{-43}\pi\delta(R_0)^3x^3$$
For $R_0$ we have $R_{S_0}=6.4\cdot10^{-43}\pi\delta(R_0)^3$. Therefore
$R_{S_x}=R_{S_0}\cdot x^3$, whence:
$$\que{R_x}{R_{S_x}}=\que{xR_0}{x^3R_{S_0}}=\que1{x^2}\cdot\que{R_0}{R_{S_0}}.$$
This implies that if $\que{R_0}{R_{S_0}}>1$; \ $\que{R_x}{R_{S_x}}\gg1$. In
the contrary case, even if the Universe had not begun as a black hole, there
would always have ben a black hole around $\omega$.

Assuming that particle $(m_e)_1$ will not ``overtake"\ others which were
initially located further from $\omega$, its velocity for very large values of
$R$ compared with $R_1$, though not necessarily when compared with $R_0$, will
be given by:
$$\left(\que{dR}{dt}\right)_1=\left (1-\que{R_{S_1}}{R_1}\right)^{1/2}=
\left[ 1-(x_1)^2\que{R_{S_0}}{R_0}\right]^{1/2}.$$

For particle $(m_e)_2$ we would have:
$$\left(\que{dR}{dt}\right)_2=\left(1-\que{R_{S_2}}{R_2}\right)^{1/2}=
\left [1-(x_2)^2\que{R_{S_0}}{R_0}\right ]^{1/2}.$$

Since $x_1<x_2$, \ $\left(\que{dR}{dt}\right)_1>\left(\que{dR}{dt}\right)_2$.
This means that the velocity of the particle which was formed more closed to
$\omega$ would tend to be greater than that of the other one. Notwithstanding
this particle would never collide with the other one formed at the same time
$t$, but on $x_2R_0>x_1R_0$. Obviously its speed when it reaches $x_2$ will be
less than $1\cdot c$, whilst the velocity of a particle $(m_e)^\prime_2$,
formed at $x_2$ just before $(m_e)_1$ reaches $x_2$ would be $1\cdot c$.
Therefore $(m_e)$, never could collide with $(m_e)_2^\prime$. On the other
hand, $(m_e)_2^\prime$ never can collide with $(m_e)_2$ formed also at $x_2$
but before $t^\prime$, and in consequence $(m_e)_1$ never could collide with
$(m_e)_2$. This is important because ``no overtaking" means the elimination
of the turbulence which would have complicated the dynamics of the particles
emerging from the Big-Bang, and which  would certainly have caused a far from
homogeneous distribution.

\section{On the increasing in the radius $R$ of the universe  and
of the arc $\ov{AB}$ between the Earth and the limits of the observable
space}

Returning to Fig. 1, we find
$$\ov{AB}=R\cdot \varphi\qquad(\varphi\ {\rm in\ radians})$$
$$\que{d\ov{AB}}{dt}=c=\varphi\que{dR}{dt}+R\que{d\varphi}{dt}\eqno{(11)}$$
$$\que{d^2\ov{AB}}{(dt)^2}=0=\varphi\que{d^2R}{(dt)^2}+R\que{d^2\varphi}{(dt)^2}+
2\que{dR}{dt}\cdot\que{d\varphi}{dt}\eqno{(12)}$$

We also know that the speed of increase of $R$, which at first was very close
to $1c$, has been slowing down because  of the gravitational pull of the
mass of the Universe $M_0$, with its centre of mass in $\omega$ (supposing an
expansion in no preferred direction, i.e. as spherical surfaces with centre at
$\omega$, starting at the Big-Bang). This means that $\que{d^2R}{(dt)^2}\leq0$
and that $\que{d\varphi}{dt}>0$, so that an increase in $\varphi$ compensates for the
decrease of $\que{dR}{dt}$ in (11) and keeps  $\que{d\ov{AB}}{dt}=1\cdot c$.

Equation (11) implies that $R\varphi=t$ (13). By substituting $t/\varphi$ for $R$ in
(10) we obtain
$$\varphi^2(At+\{\ \})^2-t^2A-t\varphi R_S=0;\eqno{(14)}$$
where
$$\{\ \}=R_0+\que{R_0}{A^{1/2}}\log\left(\que t{R_0\varphi}\right)^{1/2}
\que{A^{1/2}+(A+R_S\cdot \varphi)^{1/2}}{A^{1/2}+1}\eqno{(15)}$$

In this curve, for \ $t=0$, \ $\varphi=0$, \ and its origin is a double point with
tangents defined by $\varphi^2\{\ \}^2-t^2A-t\cdot\varphi\cdot R_S$, one has positive
gradient, $\que\varphi t=\que1{R_0}$; the other negative,
$\que\varphi t=\que{(R_S-R_0)}{R_0^2}$.

The curve has a double asymptote parallel to the axis $t=0$, given by
$(At+\{\ \})^2=0$, and two asymptotes parallel to the axis $\varphi=0$ given by
$\varphi^2A^2-A=0$; \ $\varphi=+A^{-1/2}$; \ $\varphi=-A^{-1/2}$.

\begin{figure}[h]
\centering
\resizebox{0.90\columnwidth}{!}{\includegraphics{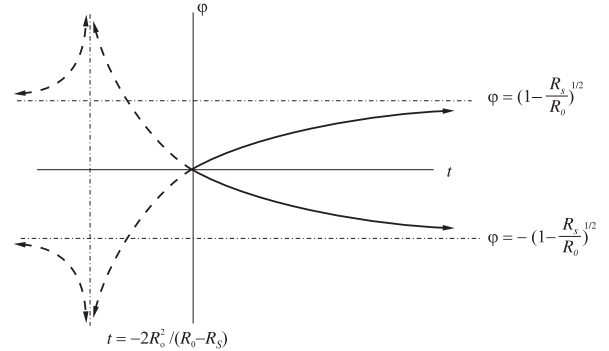}}
\caption{Path of curve \ $\varphi^2(At+\{\
\})^2-\varphi t R_S-t^2A=0$.}
\end{figure}

The valid part of this representation (the continous line which is drawed on
the region $t\geq0$;\ $\varphi\geq0$) shows that for $t=0$, \ $\varphi=0$, \
(we must remember that $t=0$, \ $R=R_0$), and that $\varphi$ increases very slowly
until after a very long time, it becames indistinguishable from the asymptote
$\varphi=\left(1-\que{R_S}{R_0}\right)^{-1/2}$.

The equations (12) and (16) would help the studies over the dynamics of the
elementary particles which  have not suffered any interaction after the
Big-Bang up to the present time. With their help we can be sure that right now
\ $\que{d\varphi}{dt}\cong0$; \
$\que{dR}{dt}\cong\left(1-\que{R_S}{R_0}\right)^{1/2}$ \ and \
$\varphi\cong\left(1-\que{R_S}{R_0}\right)^{-1/2}$.

\section{Recapitulation of definitions and relations and exploitation of the model}

\subsection{ Definitions and relations}

\begin{itemize}
\item $M_\varphi=$ Mass of the Observable Universe.

\item $M_0=$ Mass of the Universe.
For $\varphi\leq\pi$, $M_0=\que{2M_\varphi}{1-\cos\varphi}$, for
$\pi<\varphi\leq2\pi$,  $M_0=\que{M_\varphi}{2+\cos\varphi}$.

\item $R_S=$ Schwarzschild radius for \ $M_0=\que{2M_0G}{c^2}$.

\item $R_0=$ Radius of the Universe, the moment at which the formation of
elementary particles has concluded, assumed to be after 2s. have
passed since the Big-Bang. \item $t_U=$ Age of the Universe. For
$H_U=90$ Km/Mparsec$\cdot$s., \ $t_U=1.086\times 10^{10}$ years;
\ for $H_U=60$ Km/Mparsec$\cdot$s., \ $t_U=1.630\times 10^{10}$
years. \item $\ov{AB}$. The definition of $\ov{AB}$ as the
distance to the limit of the Observable Universe implies that the
length of $\ov{AB}$ in light years is equal to the age of the
Universe in years, since in Fig. 6 the valide alternative for the
representation of $\varphi$, starts from $t=0$, \ $\varphi=0$. \item
$\ov{AB}=R\cdot\varphi.$ \item
$\que{d\ov{AB}}{dt}=1=\varphi\cdot\que{dR}{dt}+ R\cdot\que{d\varphi}{dt}$;
\ whence \ $R\varphi=t$. \item
$\que{dR}{dt}=\left(1-\que{R_S}{R_0}+\que{R_S}R\right)^{1/2}$; \
see (7) in p. 10; \ $\que{dR}{dt}$ \ for \ $t=t_U$ \ is the
present speed of increase of $R$, which when $R\rightarrow\infty$ tends to
\ $\left(1-\que{R_S}{R_0}\right)^{1/2}$, the direction of the
parabolic branch in Fig. 4. \item {\em The present value of
$\varphi$}. Fig. 6 shows that $\varphi$ began as zero, at $t=0$, \
$R=R_0$, and has evolved by asymptotic approach to
$\varphi=\left(1-\que{R_S}{R_0}\right)^{-1/2}$, \ which must be
approximately equal to its present value, which is given exactly
by \ $\varphi=\que{\ov{AB}}R=\que{t_U}R$. \item {\em The value of} \
$\que{d\varphi}{dt}$, \ at present very close to zero, is obtained by
substituting, in \ $\varphi\que{dR}{dt}+R\que{d\varphi}{dt}=1$, the value
of \ $\que{dR}{dt}=\left(1-\que{R_S}{R_0}+\que{R_S}R\right)^{1/2}$
\ and is:
$$\que{d\varphi}{dt}=\que1R\left\{1-\que{t_U}R\left(1-\que{R_S}{R_0}+
\que{R_S}R\right)^{1/2}\right\}$$
\end{itemize}

\subsection{Exploitation of the model}

The evaluation of the length of $\ov{AB}$ derives from: A) Measurements of
the redshifts of the furthest cosmic objects. B) Evaluations of $H_U$,
through the observation of redshifts and other characteristics of many
cosmic objects. These measurements and evaluations lead to values of $H_U$
between 60 and 90 km/Mpsc.s. Finally, the value of $M_\varphi$ has been obtained
from a preliminary evaluation which must be improved.

The relation $t_U=\que{\ov{AB}}{H_U}$ introduces into the
evaluation of the age of the Universe the uncertainty implicit in
the evaluation of $H_U$, which means that $1.096\times10^{10}\
{\rm years}<t_U<1.626\times10^{10}$ years.

It is reasonable to suppose that $1<\varphi<\pi$ radians, and for each value of
$\varphi$ we have a value of $R$ given by $R=t_U/\varphi$, because the measure of
$\ov{AB}$ in light years is the same as the measure of $t_U$ in years. After
recalling all these facts, we can use equations (12) and
(16) to obtain sets of coherent values of $R$, $R_S$, $R_0$, $dR/dt$, $\varphi$ and
$d\varphi/dt$ for different values of both these unrelated magnitudes $H_U$ and
$\varphi$, whose fields of variation we have just settled within relatively strict
limits. For each pair of values of $H_U$ and $R_0/R_S$, there is only one value of
$R_U$ that satisfies (12), but the values of $\varphi$;
$R_S$; $R_0$ and $dR/dt$ depend only on $R_0/R_S$ and on $M_0$. Present conditions depend on what was
the value of $R_0/R_S$ when the Universe was born.
We must state here that $R_U$ is the radius which sets the boundaries for the
Universe of matter, subject to the attraction of $M_0$ placed in $\omega$. Beyond
$R_0$, we must consider $R_L=t_U\cdot c$, whose measurement in light years is
the same as $t_U$ in years and which corresponds to the distance travelled,
since the Big-Bang by the light not transformed in matter.

As it can be seen in Tables IV and V, the very low values of $R_0/R_S$ always
correspond to $\varphi\simeq\pi$, and the high values of this quotient correspond
to $\varphi$ tending to 1 radian. No value is given for $\varphi>\pi$ radians. The
value $R_0/R_S=1.00$ correspond to $\varphi\rightarrow\infty$, which is inadmissible.

Equation (11), therefore, implies the rejection of its solution for
$\que{R_0}{R_S}=1$, and the non-existence of any solutions for
$\que{R_0}{R_S}<1$, because its term
$\que{R_S}{A^{1/2}}\cdot\left(\que{R}{R_0}\right)^{1/2}
\que{A^{1/2}+(A+R_S/R)^{1/2}}{A^{1/2}+1}$ \ would include square roots of
ne\-ga\-ti\-ve numbers when $\que{R_S}{R_0}>1$, because
$A=\left(1-\que{R_S}{R_0}\right)$.

\begin{table}
\caption{Values of $\varphi$, $R_0$, $R_S$ and $dR/dt$  for different values
of $R_0/R_S$.}
\begin{ruledtabular}
\begin{tabular}{||c|c|c|c|c||}
$R_0/R_S$ & $\varphi$ & $R_S$ & $R_0$ & $dR/dt$ \\
& radians & $10^6\ {\rm l.y.}$ & $10^6\ {\rm l.y.}$ & $c$\\ \hline
1.112745 & $\pi$ & 2.216 & 2.465842 & 0.3190\\ \hline
1.20 & 2.449 & 2.504 & 3.004906 & 0.4087\\ \hline
1.35 & 1.964 & 3.204 & 4.325892 & 0.5096\\ \hline
1.50 & 1.732 & 3.819 & 5.728286 & 0.5777\\  \hline
1.75 & 1.528 & 4.632 & 8.106675 & 0.6550\\ \hline
2.00 & 1.414 & 5.251 & 10.501669 & 0.7074\\ \hline
3.00 & 1.225 & 6.707 & 20.120639 & 0.8168\\ \hline
5.00 & 1.118 & 7.878 & 39.392139 & 0.8947\\ \hline
10.00 & 1.054 & 8.759 & 87.591861 & 0.9490\\ \hline
20.00 & 1.026 & 9.200 & 184.000595 & 0.9750\\
\end{tabular}
\end{ruledtabular}
\end{table}

\begin{table*}
\caption{Values of $R_U$ for different sets of values of $H_U$ and $R_0/R_S$
(in light years).}
\begin{ruledtabular}
\begin{tabular}{||c|c|c|c||}
$R_0/R_S$ &  $H_U=\que{60\ {\rm Km}}{{\rm Mpsc} {\rm s}}$ &
$ H_U=\que{75\ {\rm Km}}{{\rm Mpsc} {\rm s}}$ &
$H_U=\que{90\ {\rm Km}}{{\rm Mpsc}\cdot{\rm s}}$ \\
& $t_U=1.6297\times10^{10}{\rm y.}$ &
$t_U=1.35808\times10^{10}{\rm y.} $ &
$t_U=1.0865\times10^{10}{\rm y.} $ \\ \hline
1.112745 & $5.213778\times10^9$ & $4.348338\times10^9$ & $3.482712\times10^9$\\ \hline
1.20 & $6.674664\times10^9$ & $5.565202\times10^9$ & $4.455608\times10^9$\\ \hline
1.35 & $8.318531\times10^9$ & $6.935039\times10^9$ & $5.551438\times10^9$ \\ \hline
1.50 & $9.430149\times10^9$ & $7.861517\times10^9$ & $6.292784\times10^9$ \\ \hline
1.75 & $1.069164\times10^{10}$ & $8.913063\times10^9$ & $7.134393\times10^9$  \\ \hline
2.00 & $1.154844\times10^{10}$ & $9.627401\times10^9$ & $7.706273\times10^9$ \\ \hline
3.00 & $1.333979\times10^{10}$ & $1.112165\times10^{10}$ & $8.903424\times10^9$ \\ \hline
5.00 & $1.462801\times10^{10}$ & $1.219821\times10^{10}$ & $9.768832\times10^9$ \\ \hline
10.00 & $1.555899\times10^{10}$ & $1.298183\times10^{10}$ & $1.0404579\times10^{10}$ \\ \hline
20.00 & $1.607763\times10^{10}$ & $1.342986\times10^{10}$ & $1.078234\times10^{10}$
\end{tabular}
\end{ruledtabular}
\end{table*}

However if ignoring (10), we consider only the evolution
of the kinetic energy given by (2), and that of $\que{dR}{dt}$ given by (5) we
could draw a curve which  crosses the axis $t=0$ at $R=R_0$, being there
tangent to $\que{dR}{dt}=1$. When $t$ grows, $dR/dt$ disminishes down to zero
at $R_t=\que{R_0}{\que{R_S}{R_0}-1}$, and from here the particle returns
towards $\omega$; when $\que{dR}{dt}$ is negative, and reaches its theoretical
maximum negative value at $R=R_0$, \ $t=\que{2R_0}{(R_S/R_0)-1}$ (which is
positive being $R_S>R_0$),

We must say here that, if $R=R_0$,
$$\log\left(\que{R}{R_0}\right)^{1/2}
\que{A^{1/2}+\left(A+\que{R_S}R\right)^{1/2}}{A^{1/2}+1}=\log 1=0$$

Figure 7 shows the hypothetical path of $R$ as a function of $t$, when
$\que{R_S}{R_0}>1$

\begin{figure}[h]
\centering
\resizebox{0.80\columnwidth}{!}{\includegraphics{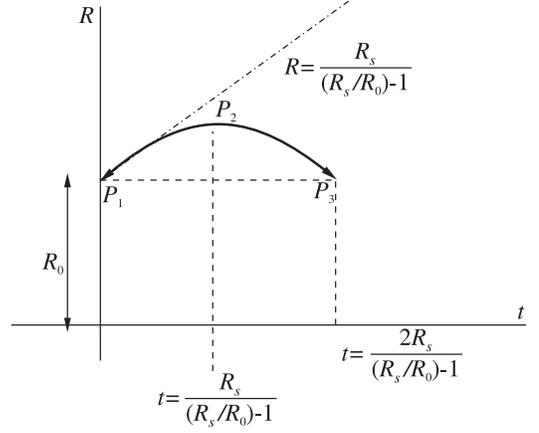}}
\caption{ Inferred path for $R$, when
$\que{R_S}{R_0}>1$.}
\end{figure}

The point $(0,R_0)$ where $dR/dt=1$ is always a point in the curve (10),
because the value of
$$\log\left(\que R{R_0}\right)^{1/2}\que{A^{1/2}+
\left(A+\que{R_S}R\right)^{1/2}}{A^{1/2}+1}$$
is zero when $R=R_0$. The other point where $R=R_0$ corresponds for
$t=\que{2R_0}{(R_S/R_0)-1}$ and is also situated on curve (10). No other values of
$t$ give real values of $R$ when introduced into (10), but we know that the
maximum $R_{\rm max}=\que{R_S}{(R_S/R_0)-1}$ corresponds to
$t=\que{R_0}{(R_S/R_0)-1}$ which agrees with (3) and that $\que{dR}{dt}=-1$ for
$t=\que{2R_0}{(R_S/R_0)-1}$, \ $R=R_0$.

Knowing these three points and the values of $dR/dt$ in each one of them we
can easily draw the path of $R$ which can be seen on Fig. 7. The only problem
is, that after reaching point $P_3$ our particle $(m_e)$ would go on towards $\omega$
at velocities greater than $c$. But $(m_e)$ would be then subject to black
hole conditions, and its behaviour might be different.

The representation of (10) is that of $R$ as function of $t$, when
$1-\que{R_S}{R_0}\geq0$. It is not valid for $1-\que{R_S}{R_0}<0$; only the
points $t=0$, \ $R=R_0$; \ $t=\que{2R_0}{(R_S/R_0)-1}$, \ $R=R_0$, \ belong to
the representation of (10) when $1-\que{R_S}{R_0}<0$. For any other value of
$t$, the equation (10) gives imaginary values of $R$. But the fact that we can
calculate easily the coordinates and the values of $dR/dt$ for points $P_1$,
$P_2$ and $P_3$ (see Fig. 7) leads us to believe that there would be a real
representation of $R$ as function of $t$ when $1-\que{R_S}{R_0}<0$, as shown
in Fig. 7. It would be convenient to know something about the possible implications
of this, and it is possible to get some useful knowledge without knowing any
equation relating $R$ to $t$ when $1-\que{R_S}{R_0}<0$.

The present value of $R_U$, given by $R_U=t_U/\varphi$, can be seen in Table IV to
be more than 1000 times the radius of Schwarzschild which corresponds to
$\varphi=1$ radian in Table I, i.e. $M_0=4.351M_\varphi$. Moreover, this value is
still increasing. On the other hand, it must be smaller than $R_{\rm
max}=\que{R_S}{(R_S/R_0)-1}$; we are still seeing the very remote cosmic objects
with a redshift and not with a blueshift.

The number $\left(1-\que{R_S}{R_0}+\que{R_S}R\right)$ must be positive, to
obtain real values for
$\que{dR}{dt}=\left(1-\que{R_S}{R_0}+\que{R_S}R\right)^{1/2}$. We have
$\left(1-\que{R_S}{R_0}\right)<0$; and $\que{R_S}R<\que{R_S}{R_0}$; because if
$R<R_0$, we would be beyond point $P_3$, and the furthest cosmic objects would
show us a blueshift instead of a redshift.

There is only one possibility left; $1-\que{R_S}{R_0}=-x$, being
$x$ very small and $\que{R_S}R>x$. Otherwise the values of $R_S$
would be greater than those of $R$, leading to values of $M_0$
absurdly greater than those which have been calculated in Table II.
For instance, if $x=1$, \ $M_0$ must be greater than
$4.351\times10^3M_\varphi$, which seems to make no sense.

If $x\leq10^{-3}$ we could have values of $R_S/R$ which make
$1-\que{R_S}{R_0}+\que{R_S}R>0,$
allowing reasonable estimates for $R_S$, \ $M_0$ and  $R_0$.

The range of variation of $x$ leads to a very short range of variation of
$\que{R_S}{R_0}$. In effect $1<\que{R_S}{R_0}<1.001$ is so short that it would
mean that the birth of the Universe was exactly determined and would exclude
any hypothesis based on chance.

\section*{REFERENCES}
[1] W. J. Kaufmann and R. A. Freedman (2000):
{\em Universe}
(Freeman, New York).

[2] P.  Appell (1942)
{\em Precis de Mecanique Rationnelle}
(Gauthier Villars, Paris).

[3] E. R. Harrisom (1986)
{\em Cosmology}
(Cambridge University Press, Cambridge).

[4] Ch. W.  Misner and K. S. Thorne (1973):
{\em Gravitation}
(Freedman, New York)
\end{document}